\documentclass[manuscript, screen=true, review=false]{acmart}

\AtBeginDocument{%
  }

\copyrightyear{2026}
\acmYear{2026}
\acmDOI{}
\acmConference[RealFab'26]{From Papers to the Real World: Making Fabrication Research Matter, a CHI'2026 Workshop}{April 14, 2026}{Barcelona, Spain}
\acmISBN{}

\usepackage{graphicx} % Required for inserting images
\usepackage{xspace}

\title{Making It Work Is the Work: Engineering Maturity as Epistemic Work}

\author{Danny Leen}
\email{danny.leen@uhasselt.be}
\affiliation{%
  \institution{Digital Future Lab, UHasselt - Flanders Make}
  \state{Diepenbeek}
  \country{Belgium}
}

\author{Stig Konings}
\email{stig.konings@uhasselt.be}
\affiliation{%
  \institution{Digital Future Lab, UHasselt - Flanders Make}
  \state{Diepenbeek}
  \country{Belgium}
}

\author{Raf Ramakers}
\email{raf.ramakers@uhasselt.be}
\affiliation{%
  \institution{Digital Future Lab, UHasselt - Flanders Make}
  \state{Diepenbeek}
  \country{Belgium}
}

\author{Kris Luyten}
\email{kris.luyten@uhasselt.be}
\affiliation{%
  \institution{Digital Future Lab, UHasselt - Flanders Make}
  \state{Diepenbeek}
  \country{Belgium}
}

\newcommand{\ilities}{\emph{Fab-ilities}\xspace}

\begin{document}
\begin{abstract}

Many HCI$\times$fabrication systems are compelling as prototypes but remain difficult to reuse, extend, or transfer beyond their original publication. A common explanation is that adoption simply takes time. We argue that the issue is more fundamental. The knowledge needed to make fabrication systems transferable, namely \emph{how they behave across different materials, machines, and users}, usually does not exist at the time of publication because the work required to generate this knowledge is rarely incentivized or rewarded. Drawing on engineering epistemology and prior debates in systems-oriented HCI, we reframe engineering maturity as epistemic work: sustained engineering effort that produces knowledge which prototyping alone cannot reveal. We propose six dimensions, \ilities, as a vocabulary to describe what aspects of fabrication artifacts have become established and what knowledge remains tacit: (1) buildability, (2) executability, (3) reliability, (4) maintainability, (5) transferability, and (6) scalability. We describe five of our own projects (JigFab, StoryStick++, Silicone Devices, LamiFold, and PaperPulse), where varied attempts at dissemination, such as commercialization, spin-offs, and market exploration, each exposed different gaps between what we published and what transfer actually required.

\end{abstract}



\maketitle

\section{Introduction}
HCI $\times$ fabrication research has become particularly good at producing compelling prototypes. We build novel tools, demonstrate new interaction techniques with physical matter, and validate them through scenarios or user studies. Yet many of these systems remain difficult to reuse, extend, or transfer beyond the original scientific contributions. This reflects a mismatch between what our community primarily \textit{values} and \textit{rewards} and what fabrication systems require to operate outside the lab. We argue that we need a more nuanced way to value ``making it work'' as part of scholarship. In today's publication culture, characteristics of good engineering work: reliability, reproducibility, maintainability, scalability, manufacturability, documentation, are structurally disincentivized. These characteristics are often treated as implementation details, secondary to the \textit{contribution} and its evaluation. As a result, engineering characteristics, though essential for the real-world utility and potential producability of the result, are often \textit{not prioritized} and dismissed during research. In fabrication research, engineering decisions are not merely implementation details. They often generate the knowledge needed to build, operate, and adapt the system beyond the initial prototype.

The sustained work of making a system function reliably produces knowledge that others need to build, run, and adapt it. This includes genuinely novel knowledge about transferability, reliability, and economic viability, but it typically emerges only through sustained practice that pushes systems beyond their first demonstration. This realization points to a structural problem. If we want systemic change in how HCI$\times$fabrication research contributes beyond publication, we must look inward and be critical about how we evaluate our papers. HCI has strong traditions of valuing concepts over execution~\cite{wobbrock_kientz_2016_contributions}. In many HCI venues, a novel idea demonstrated through a prototype, supported by a scenario and a focused study, is treated as sufficient. But if fabrication research aims for broader use---whether through replication, extension, pedagogy, partnerships, or other forms of uptake---then the engineering work that makes systems reproducible, reliable, and legible to others cannot remain an afterthought.

We argue that \textbf{engineering maturity should be valued as epistemic work}, alongside conceptual contributions. The community needs to better recognize and reward the work required toward real-world adoption, treating these engineering challenges as legitimate research outcomes. This would move the field closer to producing systems that can survive outside the lab.
We makes three contributions in this paper:
\begin{itemize}
\item[(1)] We reframe engineering maturity in HCI×fabrication as epistemic work, grounded in engineering epistemology and prior debates in systems-oriented HCI.
\item[(2)] We propose a six-dimensional vocabulary ``\ilities'', for describing the maturity of fabrication artifacts: buildability, executability, reliability, maintainability, transferability, and scalability.
\item[(3)] We reflect on five fabrication projects to illustrate recurring gaps between systems that can be published at a scientific venue, and systems that can be transferred to practice.
\end{itemize}

\section{Engineering Maturity as Epistemic Work}
Concerns that HCI ideas often fail to move beyond the laboratory are longstanding. One common interpretation is that such systems are simply premature. Buxton’s Long Nose of Innovation describes how influential ideas may take decades before reaching widespread adoption, suggesting that the path from invention to impact is inherently long~\cite{buxton_long_nose}. However, the Long Nose perspective assumes that the knowledge required for adoption already exists and only awaits the right timing or market conditions. In practice, much of the knowledge required for transfer, such as how systems behave across materials, environments, infrastructures, and user groups, does not yet exist at the time of publication. This knowledge often emerges only through sustained engineering, deployment, and iteration beyond the initial research prototype.”
In HCI$\times$fabrication, the knowledge required to make a system transferable (how it behaves across materials, machines, and users) typically does not yet exist at publication time, because the work that would generate it was never incentivized. When this work is treated as mere implementation, the knowledge it produces remains tacit. As a result, it becomes harder for the community to verify claims, compare approaches, and build cumulative knowledge. 

The philosophy of engineering has long recognized that engineering practice is itself a form of knowledge generation. Vincenti showed that engineering knowledge is not simply applied science~\cite{vincenti1990}: it includes practical knowledge from building and testing, functional knowledge about how components behave, and knowledge about what it takes to make systems work reliably. This knowledge emerges from confronting what doesn't work: from understanding constraints through building against them. HCI has grappled with related questions: technical HCI has argued that systems cannot be evaluated solely through usability studies~\cite{ledo_2018_toolkit}, and that system qualities such as flexibility, expressive leverage, and scalability should be considered evaluation criteria~\cite{olsen2007evaluating}. Community voices have articulated how review practices can discourage the submission of systems work~\cite{kaye_2009_give_up}, and Greenberg and Buxton caution that premature usability evaluation can harm early-stage innovations~\cite{greenberg_buxton_2008_harmful}. Together, these research lines motivate a reframing. In fabrication research, engineering maturity is not a postscript to the real contribution. It is often an essential part of the contribution. We therefore propose treating engineering maturity as a legitimate research outcome. It is the knowledge that enables others to build, run, and transfer a system. We treat engineering maturity as epemistic work that produces new knowledge. It is systematic engineering that creates transferable know how about how and why a system works, how it fails, and how others can adapt it. This work turns hidden lessons into shared evidence, so others can check claims, compare systems fairly, and build on prior results, instead of treating the effort as a secondary implementation detail.

\subsection{Dimensions of Engineering Maturity}
To make engineering maturity actionable, we introduce a set of \ilities as a starting vocabulary for describing maturity in fabrication systems. Specifically, we sketch six dimensions (Table~\ref{tab:maturity}) that capture what a contribution makes buildable, runnable, and transferable beyond the original lab setting. This framing is inspired by a prior workshop focusing on an Interaction Engineering Glossary and a Repository for interactive software ``\ilities''~\cite{ilities2015}, which aimed to elicit reference properties for assessing the quality of engineered HCI products. Our \ilities are not a checklist or a compliance target. They are descriptors for articulating what kind of maturity a contribution achieves, what remains unresolved, and what that maturity enables for others. We do not claim these dimensions are exhaustive or final. We offer them to make maturity discussable and reviewable, and to prompt a broader conversation about what fabrication research in HCI should value beyond novelty.
\begin{table}[hb]
\centering
\caption{The \ilities dimensions of engineering maturity for HCI$\times$fabrication systems.}
\label{tab:maturity}
\begin{tabular}{p{0.18\linewidth}p{0.74\linewidth}}
\hline
\textbf{Dimension} & \textbf{Definition and typical evidence} \\
\hline
Buildability & Others can reconstruct the artifact: bill of materials, fabrication files, tolerances, assembly instructions, and notes on substitutable parts or fragile dependencies. \\
Executability & Others can operate the system without heroic effort: installation instructions, dependencies, calibration procedures, and a ``known good'' pipeline or test case. \\
Reliability & Behavior is characterized under variability: documented failure modes, sensitivity to materials and environment, and mitigation strategies. \\
Maintainability & The artifact can survive time: modular structure, versioning, documented assumptions, and clear limitations that prevent misinterpretation. \\
Transferability & Guidance exists to adapt the system: what must remain fixed vs.\ what can change; porting notes across machines, materials, and workflows. \\
Scalability & Cost structure \& manufacturing complexity are characterized: bill of materials with pricing, and identification of cost-sensitive components affecting viability at different scales. \\
\hline
\end{tabular}
\end{table}

\emph{\textbf{Buildability}} asks whether others can physically reconstruct the artifact. Materials vary between suppliers and regions; components get discontinued within months of publication. The open-source hardware community has developed practices to address this~\cite{oshwa_definition}. Buildability requires identifying which specifications are critical, which components can be substituted, and which assembly sequences contain implicit constraints that only surface when someone else attempts reconstruction.

\emph{\textbf{Executability}} asks: can someone who did not build the system actually operate it? A system may be reconstructable yet still require its original creators, because calibration is undocumented, error recovery depends on tacit knowledge, or the workflow assumes internalized expertise. This spans software (environment setup, dependency conflicts, implicit build assumptions) and physical operation (material placement, focus adjustment, speed/power tuning). The two interact: fabrication systems involve software controlling hardware, and failures can occur anywhere in that chain.

\emph{\textbf{Reliability}} concerns behavior under real-world variability, for example: humidity affects adhesion, temperature influences curing, tool wear accumulates. Characterizing what fails first under what conditions requires sustained observation exceeding typical paper timelines. Recent work on FEDT~\cite{savage2025fedt} highlights that such characterization experiments are performed inconsistently, tools that support systematic experimental design offer one path forward.

\emph{\textbf{Maintainability}} asks whether the system can be maintained without its original authors. Software engineering treats maintainability as first-class~\cite{fowler1999refactoring}; fabrication research does not. Yet physical systems face additional challenges: components wear out, materials degrade, and suppliers discontinue parts. A system that cannot be maintained becomes frozen at publication and is unable to evolve or adapt when dependencies change.

\emph{\textbf{Transferability}} asks whether the system can travel to new contexts—different machines, materials, or users—and what must remain fixed. This is distinct from robustness: a system may be robust locally yet fail completely elsewhere. Replication studies show how contextual factors invisible to authors determine success in new settings~\cite{wilson2014replichi}. Answering transferability questions requires systematic experimentation with variations.

\emph{\textbf{Scalability}} confronts economic realities that prototypes leave unexamined. What does it cost to build ten versus a hundred? Research prototypes use components chosen for convenience—hand-soldered connections, expensive development boards, specialty materials—obscuring whether the contribution could ever leave the lab. Understanding which cost drivers dominate at volume, which steps need automation, and which choices trade convenience against viability is itself valuable knowledge.

\section{Lessons from Disseminating HCI$\times$Fabrication Systems}
We pursued multiple dissemination pathways for our fabrication research projects. In each case, we found gaps between what we had published and what was needed to move to the next stage. All projects listed below produced functional prototypes that were used beyond the specific demonstrations reported in their publications.

\subsection{Cases of valorisation attempts}
\textbf{JigFab}~\cite{leen_2019_jigfab} was developed in close collaboration with Fablab Genk, an open access makerspace run by one of the authors. We worked with woodworkers, hobbyists, and engineering students who use the lab daily. The system helps users design and fabricate custom woodworking jigs using digital fabrication tools. When we pursued commercialization, we filed a patent application and engaged an external marketing agency to identify potential industrial markets, but we found little traction. Prospective commercial users wanted a more mature system. They asked for clearer failure modes, stronger documentation, and behavior that remained reliable across different machines, materials, and workflows. In addition, the project involved multiple universities, which made intellectual property valorization more complex. Shared ownership introduced additional coordination, legal constraints, and procedural barriers that effectively raised the threshold for moving from a research prototype to a commercial product.
\begin{figure}
    \centering
    \includegraphics[width=0.7\linewidth]{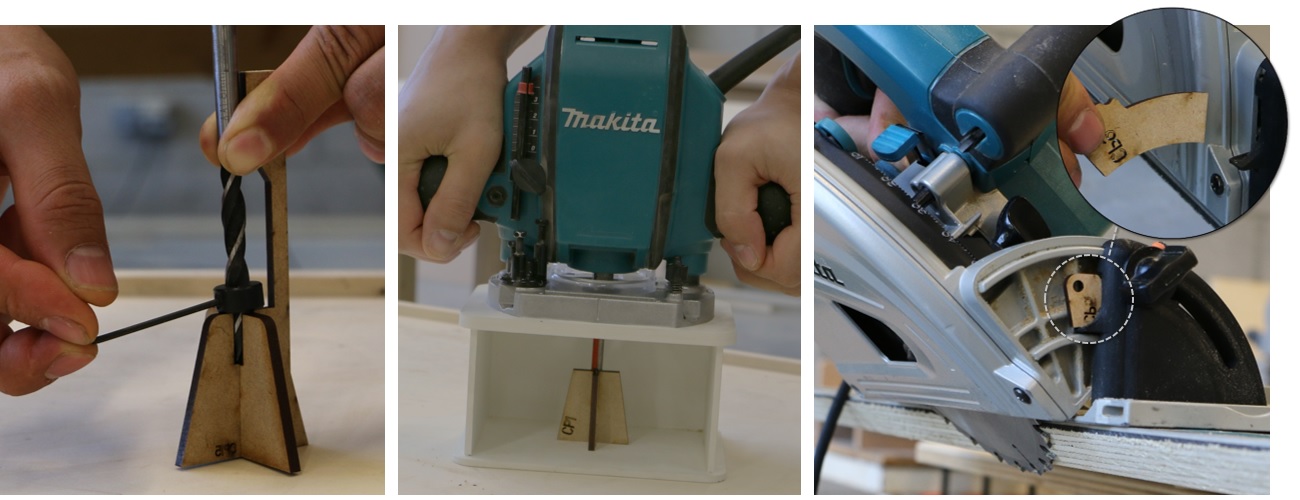}
    \caption{\textbf{JigFab} helps makers design and fabricate custom woodworking jigs. Commercialization exposed gaps in robustness, documentation, cross-context reliability, and multi-institution IP complexity.}
    \label{fig:jigfab}
\end{figure}

\textbf{StoryStick++}\cite{konings_2026_storystick} is a phone clip on that guides measuring and marking with minimal reliance on numbers, units, or mental arithmetic. This reduces common errors. StoryStick++ is currently in market research. We repeatedly see the same pattern. Companies understand what the system does, but they struggle to imagine a finished product based on the bare bones prototype we can provide. The gap is therefore not conceptual. The gap is practical and economic. The knowledge needed to build, run, and adapt the system to their context is not written down and remains in the heads of the developers. As a result, basic questions about cost structure, manufacturing complexity, and market viability remain unanswered.
\begin{figure}
    \centering
    \includegraphics[width=0.6\linewidth]{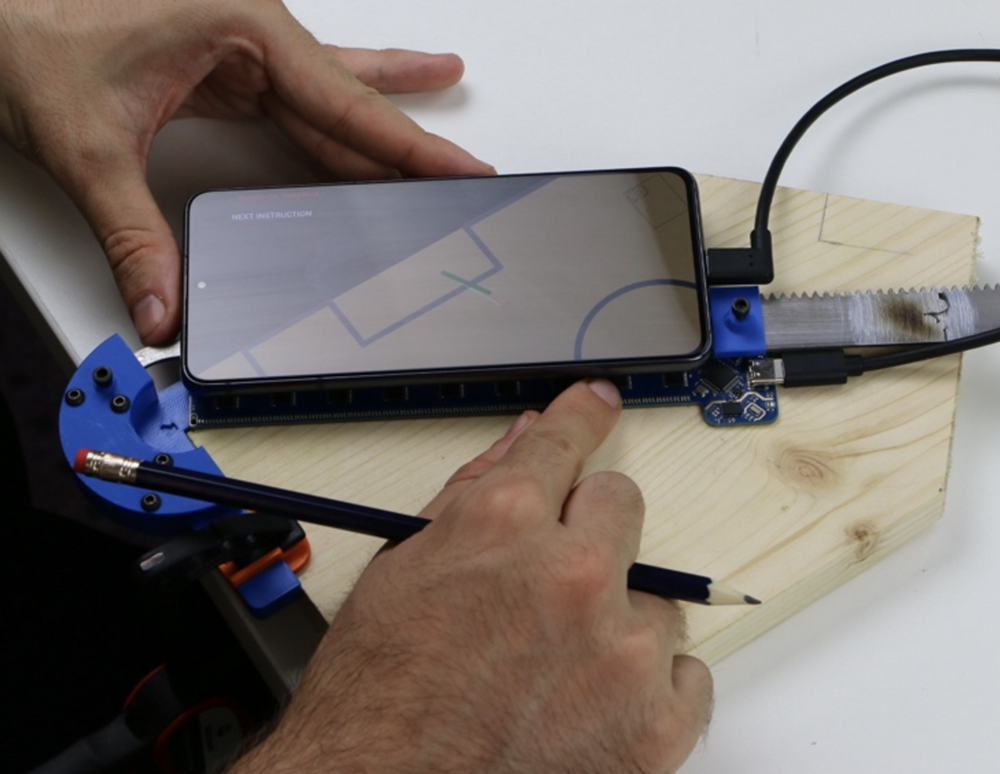}
    \caption{\textbf{StoryStick++} is a phone clip on for measuring and marking without mental arithmetic. Market research highlighted tacit build knowledge, unclear manufacturing effort, and missing cost and viability answers.}
    \label{fig:storystick}
\end{figure}

\textbf{Silicone Devices}\cite{nagels2018silicone} presents a DIY workflow that uses a CO\textsubscript{2} laser cutter to fabricate stretchable electronic devices in cast silicone with embedded liquid metal traces, computation, and sensors. To support replication, we published a step by step tutorial as an Instructable~\cite{instructable_silicone_devices}. In practice, the workflow requires careful execution, so workshops at our university and at the University of Chicago were facilitated by one of the authors. The project also received national valorization funding to attempt to set up a spin-off company on stretcheable electronics. We received funding for a two year trajectory to translate the technique into an industrial process. During that trajectory, the focus shifted from the original HCI goal of democratizing fabrication toward industrial applications, where it faced strong competition from established manufacturing techniques. The track ended when the lead researcher moved into a different role. The work, however, continues to inform research in both materials science and HCI.
\begin{figure}
    \centering
    \includegraphics[width=0.7\linewidth]{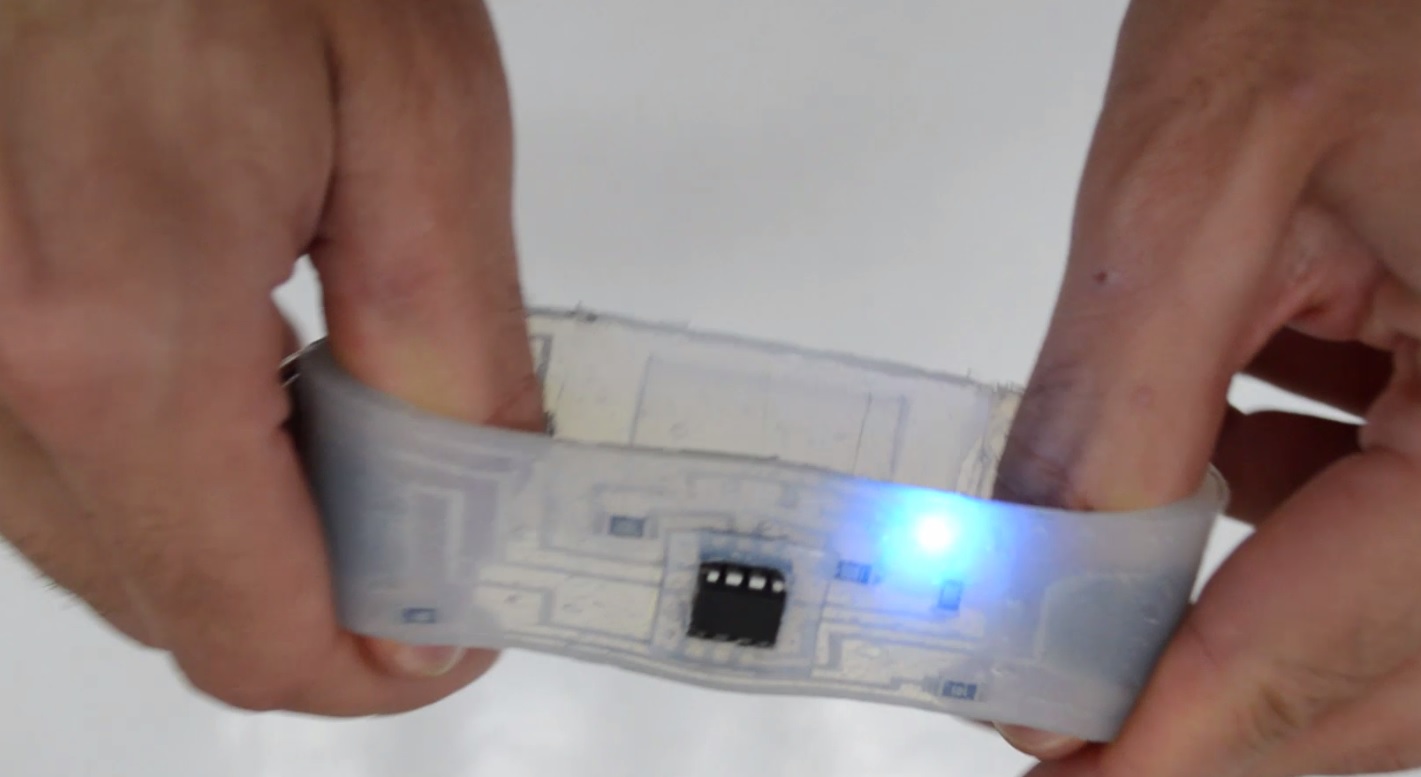}
    \caption{\textbf{Silicone Devices} enables DIY stretchable electronics in cast silicone using a CO\textsubscript{2} laser cutter and liquid metal. Transfer required process control and industrialization knowledge beyond the published workflow.}
    \label{fig:siliconedevices}
\end{figure}

\textbf{PaperPulse}~\cite{ramakers_2015_paperpulse,siggraph_RamakersTL15a} enabled non-experts to produce interactive paper devices using printed conductive traces. We explored commercialization together with a professional partner and investigated patenting options, including prior-art checks with an external intellectual property firm. However, the upfront costs and the uncertainty about a viable market path made it difficult to justify continuing the trajectory. The core novelty of PaperPulse aligns naturally with consumer oriented (B2C) markets, which are particularly difficult to sustain in Europe. This experience exposed a broader lesson. Many HCI system contributions are designed around specific user groups and contexts of use. When they are repositioned for different domains, such as expert, education or industrial users, the original contribution often loses relevance. This contrasts with general purpose techniques in materials science, in this case printed electronics, that tend to transfer more easily across application areas.
\begin{figure}
    \centering
    \includegraphics[width=0.7\linewidth]{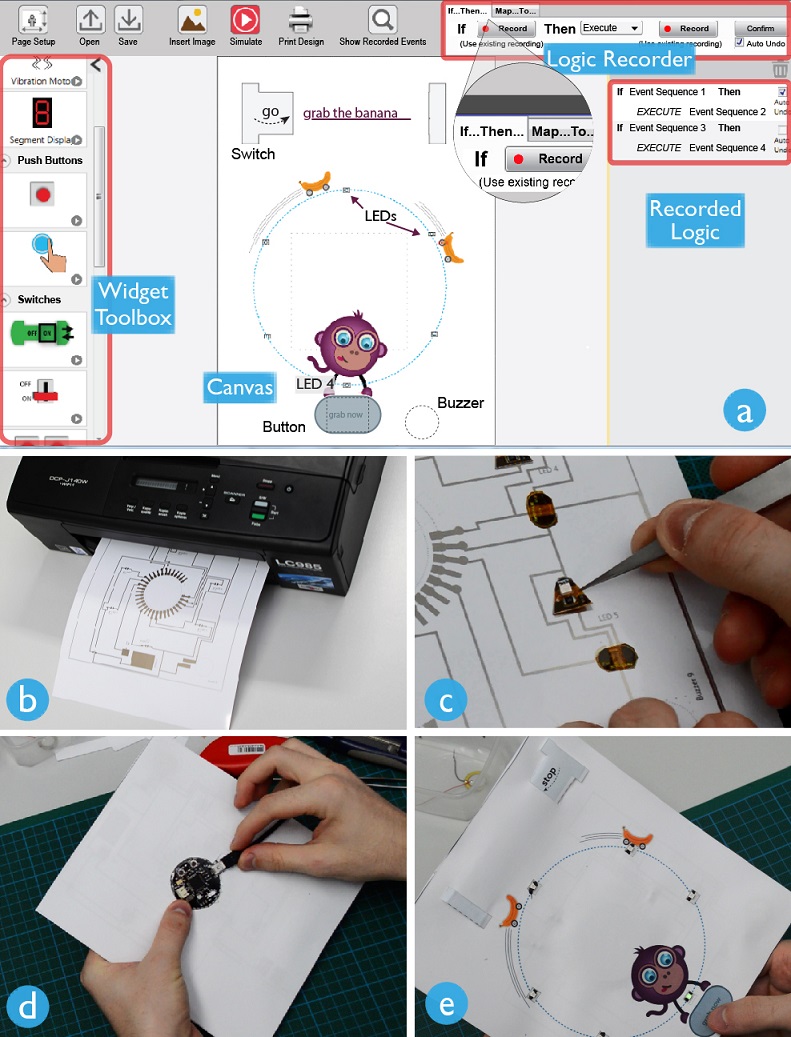}
    \caption{\textbf{PaperPulse} enables interactive paper devices with printed conductive traces. Valorization attempts surfaced high upfront costs, market uncertainty, and limited transfer when repositioning a user-specific HCI contribution.}
    \label{fig:paperpulse}
\end{figure}

\textbf{LamiFold}~\cite{leen_2020_lamifold} exposed a different facet of the maturity gap, namely reproducibility. LamiFold’s lamination workflow selectively cuts and glues stacks of sheet material to embed functional mechanisms. In practice, the process is highly sensitive to machine and material parameters. Laser power, cutting speed, kerf width, and material thickness interact in ways that are hard to describe in general terms. Calibration settings that yield reliable results on one laser cutter often fail on another. As a result, the knowledge required to make LamiFold work reliably across different setups remained largely undocumented and stayed tied to hands on experience rather than transferable instructions.
\begin{figure}
    \centering
    \includegraphics[width=0.5\linewidth]{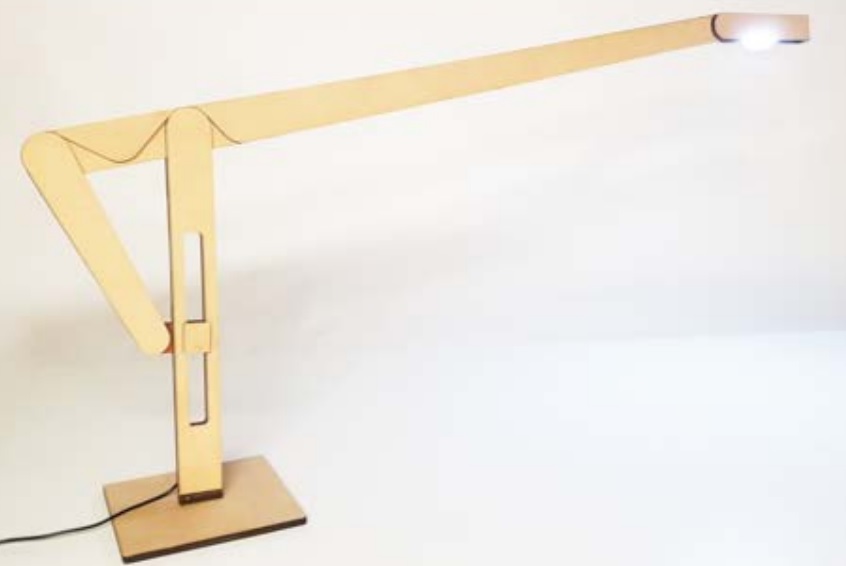}
    \caption{\textbf{LamiFold} embeds mechanisms by cutting and laminating stacked sheets, resulting in zero-effort assembly. Reproducibility is fragile because performance depends on tightly coupled machine and material parameters.}
    \label{fig:lamifold}
\end{figure}

\subsection{Barriers to Uptake Beyond Publication}
One pathway has worked consistently. Our projects have repeatedly influenced the classes we teach on designing and building interactive systems, as well as fabrication and prototyping, so the research spreads through education. The systems are complex, broad in scope, and deep in technical detail, which makes them difficult to master and therefore well suited as training material for students. However, that same complexity is also a barrier to use outside the lab. What is acceptable as open ended exploration in research and education becomes hard to operationalize in practice, where users expect stable behavior, clear procedures, and predictable costs.

When we pursued other pathways such as commercialization, licensing, spin offs, and other types of market adoption, we encountered various barriers that prevented uptake. The reasons differed across projects, but the pattern was similar. Reliability was not characterized, operational know-how remained undocumented, cost-benefit models were missing, designs were tightly bound to a specific (lab) context and did not transfer well to real life usage, and the documentation needed for others to continue the work was often absent.

\section{Conclusion}
HCI$\times$fabrication has developed a comfortable hypocrisy. We regret that our prototypes do not reach the real world, yet we systematically devalue the very work that would make them travel. We publish the demo and call it done. We move on to the next novelty, only to wonder why nothing sticks. The engineering work we dismiss as ``mere implementation'' is not overhead — it is where much of the knowledge lives. What breaks, what varies, what costs, what transfers. This is precisely what we and others need to build on or disseminate our contributions.  We argue that engineering maturity is epistemic work, not implementation overhead. 

We proposed six \ilities dimensions: buildability, executability, reliability, maintainability, transferability, and scalability. The \ilities serve as a vocabulary for articulating what fabrication artifacts have consolidated and what remains tacit. We included five illustrative projects that show a consistent pattern. Once we moved beyond publication and tried to translate the work into sustained use, we encountered barriers that the paper contributions had not resolved. If we are serious about real-world impact, and not only about stating that we value it, then the field must change what it rewards. Engineering maturity is epistemic, knowledge-producing work. It is time we evaluate and credit it as such.

\section*{Acknowledgements}
This was was partially supported by Flanders Make, the strategic research center for the manufacturing industry in Flanders, and by the Flemish Government under the ``Onderzoeksprogramma Artificiële Intelligentie (AI) Vlaanderen'' program, R-13509. We like to thank Mieke Haesen and the Tech Transfer Office of UHasselt for their continuous support.

\bibliographystyle{ACM-Reference-Format}
\bibliography{references}

 \end{document}